\documentclass[prl,reprint,aps,superscriptaddress,longbibliography]{revtex4-2}

\usepackage{amsmath}
\usepackage{amssymb}
\usepackage[varbb,varg]{newtx} 
\usepackage{microtype}
\usepackage{graphicx}
\usepackage{dcolumn}
\usepackage{bm}
\usepackage{braket}
\usepackage{feynmp}
\usepackage{graphics}
\usepackage{enumerate}
\usepackage[dvipsnames]{xcolor}

\usepackage{comment}
\usepackage[normalem]{ulem} 
\usepackage{hyperref}
\hypersetup{colorlinks=true, linkcolor=NavyBlue, urlcolor=NavyBlue, citecolor=NavyBlue} %
\usepackage[capitalise]{cleveref}

\renewcommand{\arraystretch}{1.5}
\everymath{\displaystyle}

\begin{document}

\title{Optimizing Chern Bands for Moore-Read States: A Parent Hamiltonian Approach}

\author{Bishoy M. Kousa}
\affiliation{Department of Physics, The University of Texas at Austin, Austin, Texas 78712, USA}

\author{Yung-Chun Chao}
\author{Jung-Jung Su}
\affiliation{Department of Electrophysics, National Yang Ming Chiao Tung University, Hsinchu 300, Taiwan}

\author{Allan H. MacDonald}
\affiliation{Department of Physics, The University of Texas at Austin, Austin, Texas 78712, USA}

\date{\today}

\begin{abstract}
The recent realization of fractional quantum anomalous Hall states in moiré materials motivates
efforts to stabilize non-Abelian topological order in the absence of a magnetic field. 
Here, we propose a framework for optimizing interactions to favor the non-Abelian Moore-Read state. 
Our approach compares band-projected three-particle spectra to that of the Moore-Read parent Hamiltonian, which is defined by a three-body interaction, and quantifies 
the comparison by defining a cost function that estimates proximity to the ideal Moore-Read limit.  
We apply our approach to Landau levels and to the Chern bands of twisted MoTe$_2$. 
\end{abstract}

\maketitle

\emph{Introduction ---}
Two-dimensional (2D) electron systems that combine strong correlations and non-trivial band 
topology can support exotic new phases of quantum matter. 
The fractional quantum Hall (FQH) effect, in which strong interactions within partially filled Landau 
levels (LLs) give rise to topologically ordered states with fractionalized excitations, is the paradigmatic 
example. FQH states in LLs are of interest both from a fundamental perspective and for potential
quantum computing applications \cite{kitaev2003fault,sarma2005topologically,nayak2008non}, and 
can be viewed as a specific example of fractional Chern insulator states (FCIs)    \cite{Neupert_FCI,DasSarma_Sun_FCI,Wen_FCI,Sheng_FCI,Bernevig_Regnault_FCI,Bernevig_Regnault_FCI_2}, 
understood to occur in systems with narrow Chern bands.  

Recently, non-LL FCIs have been realized in twisted MoTe$_2$ homobilayers
\cite{FCI_Experiment1,FCI_Experiment2,FCI_Transport1,FCI_Transport2,FengchengTopology} and in 
rhombohedral graphene multilayers aligned with hexagonal boron nitride (hBN) \cite{lu2024fractional}. 
These developments raise two broad classes of questions. The first concerns
qualitative differences between FCIs and their conventional FQH counterparts. 
For example, the reduction from continuous to discrete translational 
symmetry due to the underlying moiré potential leads to dispersion of anyonic excitations \cite{eslam_dispersing_anyons,mora_anyons} and qualitatively modifies optical response
properties related to neutral collective modes \cite{Fengcheng_MoireAssisted,rotons,liang_shining,Toby_Structurefactor}.
The second class concerns potential exploitation of the greater tunability of moiré systems 
to realize more robust fractionalized topological phases in more experimentally accessible regimes. 
This potential is particularly appealing for non-Abelian topological states, which are
the focus of the present work.

The braiding statistics of non-Abelian topological states is a potential resource 
for topological quantum computation \cite{kitaev2003fault,sarma2005topologically,nayak2008non}.
The possibility of realizing non-Abelian states in moiré FCIs has therefore 
attracted considerable attention \cite{higher_vortexability,reddy2024non,ahn2024non,chen2025robust,di_xiao_non,fengcheng_generalized,xu2025multiple,jie_wang_non, emil_nonAbelian}.  
In conventional FQH systems, the half-filled $n=1$ LL appears to support robust non-Abelian FQH states \cite{willett1987observation,MR,5/2_review}.  
Consequently, the usual approach to searching for zero-field non-Abelian states 
has been to identify systems whose Chern bands best approximate the $n=1$ LL, 
appealing to the close connection between moiré Chern bands and LL physics \cite{ledwith_blg,jie_wang_exact_LLL,AdiabaticAproximation1,AdiabaticAproximation2}.
There are indications, however, that close LL mimicry is not necessary for stabilization of fractionalized topological phases. For instance, recent studies suggest that FQH gaps may be enhanced away from the ideal LL limit \cite{kai_sun_dan_parker_beyond_LLL,sarkar2026similar}, and 
that fractionalized topological order can even arise in topologically trivial 
bands \cite{emil_trivial,newton_fci_c0}. These observations motivate searching for non-Abelian FCIs without requiring proximity to a particular Landau level, instead developing alternate criteria that target the desired non-Abelian state more directly. 

In this work, we propose a criterion for optimizing band-projected interactions to favor the non-Abelian 
Moore–Read state that goes beyond $n=1$ LL mimicry. Our approach compares the three-particle spectrum of a 
projected interaction directly to that of the Moore-Read parent Hamiltonian by using a three-body 
generalization of Haldane pseudopotentials and defining a cost function that 
quantifies proximity to the Moore-Read parent Hamiltonian. In LLs, we obtain the three-particle spectrum 
analytically in terms of the two-body Haldane pseudopotentials and benchmark the cost function using 
Coulomb interactions projected to LLs with different orbital characters. 
We then use the cost criterion as a design principle to engineer two-
body interactions favorable to the Moore-Read state 
and validate the resulting interactions using many-body exact diagonalization (ED), finding spectra consistent with Moore–Read physics.
We extend the framework to Chern bands and apply it to twisted MoTe$_2$ with the goal of identifying experimentally relevant parameter regimes favorable for stabilizing non-Abelian FCIs in the absence of a magnetic field.

\textit{Haldane pseudopotentials and parent Hamiltonians ---} 
Rotationally invariant pairwise interactions in LLs can be decomposed into a sum 
of projectors onto two-body relative angular momentum states, $\hat{V} = \sum\nolimits_m v_m  |m\rangle\langle m|$,
where $m$ is odd for spin polarized fermions and the coefficients $v_m$ are the Haldane pseudopotentials. 
The fermionic Laughlin state is an exact ground state of a parent Hamiltonian in which only $v_1 \ne 0$.
The Moore-Read state is analogously the exact ground state of a parent Hamiltonian involving a three-body interaction, as detailed below. In the 
following we use the complete orthonormal basis for the fermionic three-body relative-motion Hilbert 
space provided by Laughlin \cite{laughlin_3body}:
\begin{equation}
\begin{aligned}
    \ket{\mu,\nu}= \mathcal{N}_{\mu\nu}&[\frac{(z_a+iz_b)^{3\mu}-(z_a-i z_b)^{3\mu}}{2i}] (z_a^2+z_b^2) ^\nu \\
    &\times e^{-(|z_a|^2+|z_b|^2)/4}
    , \\
        \mathcal{N}_{\mu\nu}=&\frac{1}{[2^{6\mu+4\nu+1} (3\mu+\nu)! \nu!\pi^2]^{1/2}},
\label{eq:three_body_basis}
\end{aligned}
\end{equation}
where $z_a$ and $z_b$ are the three-body relative-motion Jacobi coordinates
\begin{equation}
    \begin{aligned}
        z_a&=\sqrt{\frac{2}{3}}\left(\frac{z_1+z_2}{2}-z_3\right),\\
        z_b&=\frac{1}{\sqrt{2}}\left(z_1-z_2\right),
    \end{aligned}
    \label{jacobi_coordinates}
\end{equation}
$\mu=1,2..$, and $\nu=0,1,2,..$. The three-body relative angular-momentum, $L=3\mu+2\nu$. 
Note that there may be more than one $|\mu,\nu \rangle$ state for a given $L$, with  
the first instance of 
non-unit-dimension angular-momentum subspaces occurring at $L=9$. 
For one dimensional $L$ sectors, the three-body pseudopotential is 
simply the coefficient multiplying the 
triplet projector $\hat{\Pi}^{(3)}_{L}= |L\rangle \langle L |$. 
For higher-dimension $L$ sectors, the most general rotationally invariant three-body interactions are represented by Hermitian matrices \cite{simon2007pseudopotentials}.
At filling factor $\nu=1/2$, the fermionic Moore-Read Pfaffian,
\begin{equation}
    \Psi_{\mathrm{MR}}=\mathrm{Pf}(\frac{1}{z_i-z_j}) \prod_{i<j} (z_i-z_j)^2
\end{equation}
is the zero-energy eigenstate of a positive-semi-definite parent Hamiltonian that is the sum over 
triples of the projector onto the lowest angular momentum $L=3$ relative motion state \cite{read_rezayi_1996, rezayi2000incompressible}:
\begin{equation}
    H_{\rm MR}= \lambda \sum_{i<j<k}\hat \Pi^{(3)}_{L=3}(i,j,k),
\end{equation}
where $\lambda>0$ is a coupling constant.  The Moore-Read state is the highest-density state 
in which electron 
triplets completely avoid occupation of the $L=3$ sector \cite{W_js_2005}. 
\begin{table}[t]
\centering
\renewcommand{\arraystretch}{2.1}
\begin{tabular}{c c l}
\hline\hline
$(\mu,\nu)$ & $L=3\mu+2\nu$ & $H_{L}=\langle \mu\nu|V_{12}|\mu\nu\rangle$ \\
\hline
$(1,0)$ & $3$ & $\frac{3}{4} v_1 + \frac{1}{4} v_3$ \\
$(1,1)$ & $5$ & $\frac{9}{16} v_1 + \frac{1}{8} v_3 + \frac{5}{16} v_5$ \\
$(2,0)$ & $6$ & $\frac{3}{16} v_1 + \frac{5}{8} v_3 + \frac{3}{16} v_5$ \\
$(1,2)$ & $7$ & $\frac{27}{64} v_1 + \frac{15}{64} v_3 + \frac{1}{64} v_5 + \frac{21}{64} v_7$ \\
$(2,1)$ & $8$ & $\frac{9}{32} v_1 + \frac{7}{32} v_3 + \frac{7}{32} v_5 + \frac{9}{32} v_7$ \\
\hline\hline
\end{tabular}
\caption{
Decomposition of the three-body energies $H_L$ into two-body Haldane pseudopotentials $v_m$ for the lowest $L$ three-body relative angular momentum states.}
\label{table: three_body_decomposition} 
\end{table}

\textit{From two-body interactions to three-particle spectra ---} 
Since real systems typically have pairwise Coulomb interactions, the practical goal 
is to identify two-body interactions that favor Moore-Read three-particle correlations. 
We therefore reorganize the two-body interaction into a sum over triplets, 
\begin{equation} 
\sum_{i<j}V_{ij}=\frac{1}{N-2} \sum_{{i<j<k}}(V_{ij}+V_{ik}+V_{jk})=\frac{1}{N-2} \sum_{{i<j<k}}\tilde{V}_{ijk}.
\end{equation} 
To characterize how an isotropic two-body interaction acts within the three-particle Hilbert space,
we reexpress the three-particle Hamiltonian in terms of Haldane pseudopotentials $v_m$: 
\begin{equation}
\begin{aligned}
H_L=\langle \mu \nu |V_{12}|\mu'\nu'\rangle= \sum_m v_m P_m(\mu, \nu; \mu',\nu')
\end{aligned}
\end{equation}
where $L=3\mu+2\nu$.  By permutation symmetry, the matrix elements of the full triplet interaction 
$\widetilde V_{123}$ differ from that of a single 
pair by a factor of three. 
Using the three-particle basis in Eq.~\eqref{eq:three_body_basis}, we find that:
\begin{equation} 
\begin{aligned} 
& P_m(\mu,\nu; \mu',\nu')= \delta_{L,L'} \chi^m_{\mu\nu} \chi^m_{\mu'\nu'} ,\\  
&\chi^m_{\mu\nu}  \equiv \left [\frac{m! (L-m)!}{2^{L-1} (3\mu+\nu)! \nu!}  \right ]^{1/2} C^m_{\mu,\nu}, \\  
&C_{\mu \nu}^m\equiv \sum_{j=0}^\nu (-1)^{\frac{m-1}{2}-j} \binom{3\mu}{m-2j} \binom{\nu}{j}. 
\end{aligned} 
\label{eq:mapping}
\end{equation} 
A derivation of this expression is described in the Supplemental Material (SM) \cite{SM}
and explicit expressions for $U_{L}$, the eigenvalues of $H_L$, in terms of $v_m$ for 
some sectors are given in Table \ref{table: three_body_decomposition}.  
These eigenvalues will be used below to construct our Moore--Read cost function.
For the entries in this Table, the angular-momentum $L$ subspace has dimension one and  
the $P_m$ values represent the probability distribution over two-body relative angular 
momentum states of any pair in a given three-body state. 

\begin{figure}
    \centering   
\includegraphics[width=0.49\textwidth]{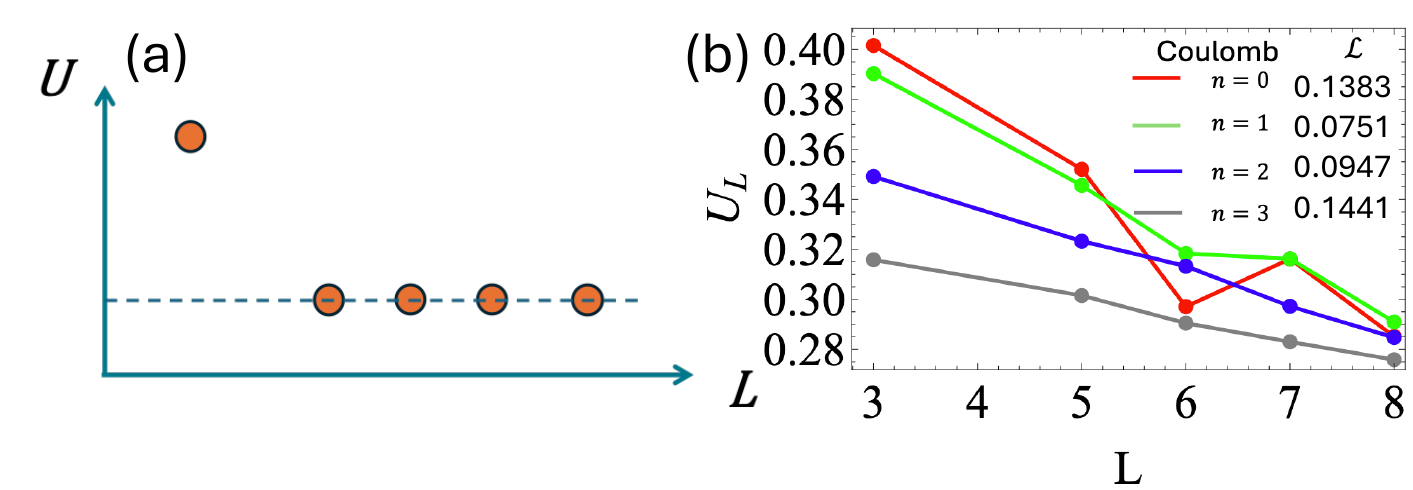}
    \caption{(a) Schematic showing the ideal three-particle spectrum of the Moore-Read parent Hamiltonian. The lowest relative angular momentum channel, $L=3$ is separated from the higher $L$ channels, which contribute only an overall energy shift when their pseudopotentials 
    have equal value. (b) Three-particle spectrum, $U_L$, of the Coulomb interaction projected to different Landau levels. The legend shows the corresponding values of the Moore-Read cost function, $\mathcal{L}$ defined in the main text. The $n=1$ Landau level has the lowest cost, consistent with its established association with non-Abelian FQH physics.}  
    \label{fig:CoulombLLs}
\end{figure}

The ground state of any three-body Hamiltonian is invariant under rescaling or $L$-independent 
shifts of the three-particle pseudopotentials.  The Moore-Read ground state is generated
when $U_3$ is the largest pseudopotential and the $U_L$ values in other sectors are equal, as illustrated in Fig.~\ref{fig:CoulombLLs}(a). We therefore seek interactions with this 
structure.  For example, one cost function that quantifies deviations from the ideal Moore-Read form is 
\begin{equation}
    \mathcal{L}= \frac{1}{\mathcal{N}}\sum_{L>3,i}^{L_{max}} w_L \left(\frac{U_{L,i}-\bar{U}}{U_3-\bar{U}}\right)^2
\label{eq: cost_function}
\end{equation}
where 
$\mathcal{N}=\sum\nolimits_{L>3}^{L_{\max}} w_L d_L$,
$i$ labels energies within a multidimensional $L$ sector, $w_L$ is a weight assigned to angular momentum channel $L$, 
and $\bar{U}$ is the weighted average across the higher $L$. 
In Eq.~\ref{eq: cost_function} the numerator penalizes variations within the higher angular momentum channels, and the denominator favors separation of $U_3$. 
Since lower relative-angular-momentum channels probe shorter-range correlations relevant to the clustering physics of the Moore-Read state, their contributions are expected to be more important, so the 
assigned weights should decrease with $L$. 
In \cref{fig:CoulombLLs}(b), we plot $U_L$ values up to $L=8$ for the Coulomb interaction projected onto the $n=0,1,2,3$ LLs and in the inset the corresponding cost functions calculated with 
$L_{max}=8$ and  $w_L=e^{-(L-3)}$.  We have confirmed that for cutoffs 
$6<L_{max}< 10$, 
any reasonable choice \cite{SM} for $w_L$ correctly identifies the $n=1$ LL as 
most favorable for Moore-Read physics \cite{willett1987observation,MR,5/2_review}.   
In the SM \cite{SM} we study how $\mathcal{L}$ changes with variations of $v_1$ and $v_3$ relative to $n=1$ Coulomb values. We find a ridge of low-cost that is qualitatively consistent with 
the line of enhanced Moore–Read overlap found in previous many-body ED studies \cite{peterson2008orbital,storni2010fractional}.
Given a Hamiltonian $H[p]$ parameterized by physical parameters $p$, the cost function in Eq.~\eqref{eq: cost_function} can provide an objective function for gradient-based Moore-Read-correlation parameter-space searches.  Evaluating the cost function requires only three-particle ED calculations, enabling efficient numerical searches.

\textit{Inverse design of two-body interactions ---}We now consider the inverse problem of designing a two-body interaction that reproduces the Moore–Read three-particle spectrum.  
The normalized Moore–Read parent spectrum ($U_3=1$ and $U_{L>3,i}=0$) can be reproduced exactly if we use $L=7$ 
as the cut-off for both two and three body pseudoptentials.
At $L=8$, however, no additional two-body pseudopotential enters, so that it is impossible to set $U_8=0$, as can be seen from Table \ref{table: three_body_decomposition}. 
The mismatch becomes increasingly severe at larger angular momentum cut-offs: the number of three-particle relative-motion states below cutoff $L_{max}$ grows asymptotically as $L_{max}^2/12$, while the number of available fermionic two-body pseudopotentials grows only linearly with $L_{max}$.
Since the Moore-Read parent Hamiltonian cannot be realized exactly in a system with 
pairwise interactions, the goal is to approximate it as closely as possible with physically
realistic interactions.

\begin{figure}
    \centering   
\includegraphics[width=0.49\textwidth]{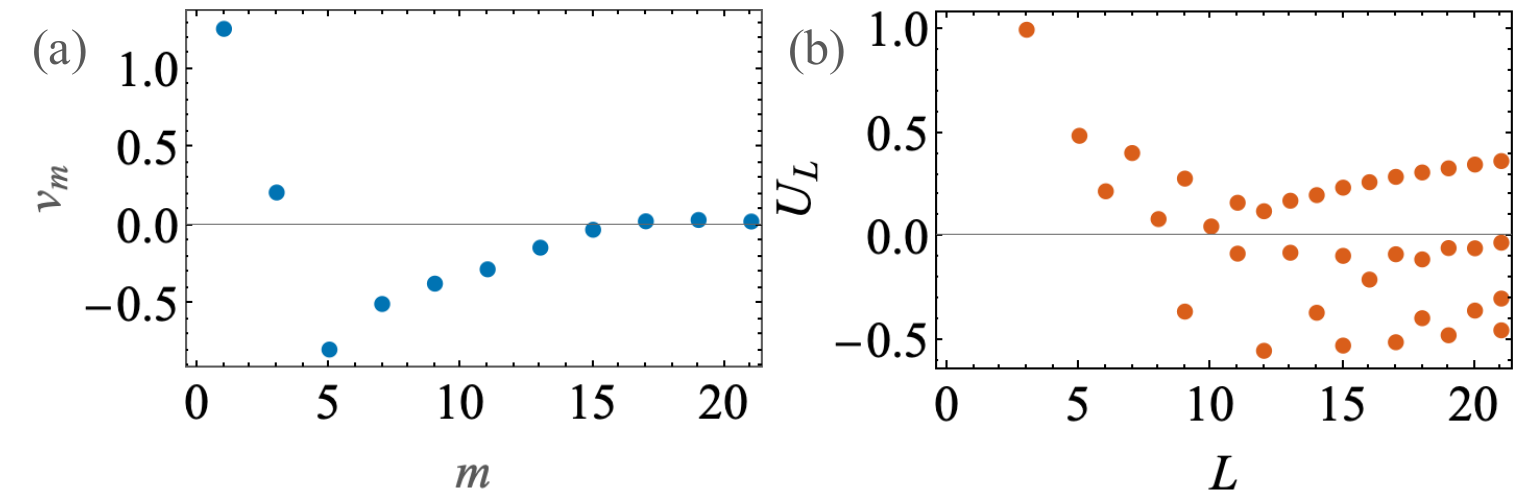}
    \caption{(a) Two-body Haldane pseudopotentials, $v_m$ of the engineered interaction obtained by optimizing the pair interaction using the parameterization of Eq. \eqref{eq:paramterized_interaction}. (b) The corresponding three-particle spectrum. The cost
    function separates the $L=3$ channel while penalizing variations among the higher $L$ channels.
    }
    \label{fig:engineered_interaction}
\end{figure}

The $m$'th Haldane pseudopotential in LL $n$,
\begin{equation}
    v_{n,m}=\frac{1}{2\pi}\int dq q v(q) [L_n(q^2/2)]^2L_m(q^2)e^{-q^2},
\label{eq:Halane_psuedo}
\end{equation}
where $q$ is measured in units of inverse magnetic length $l_B^{-1}$. 
Since physically realistic $v(q)$ vanish at large $q$, we parameterize the interaction as
\begin{equation}
    v_\sigma(q)=\sum_\alpha c_\alpha q^{2\alpha} e^{- q^2/2\sigma^2}
\label{eq:paramterized_interaction}
\end{equation}
Combining Eq.~\eqref{eq:paramterized_interaction} with \eqref{eq:mapping} gives
\begin{equation}
\begin{aligned}    &H_{\mu',\nu';\mu,\nu}=\delta_{L,L'}\sum_\alpha T_\alpha^{(n)}(\mu, \nu,\mu',\nu') c_\alpha,\\
&T_\alpha^{(n)}(\mu, \nu,\mu',\nu') =\sum_{m}I_{m,\alpha} ^{(n)}P_m(\mu,\nu, \mu',\nu'), \\
&I_{m,\alpha}^{(n)}= \frac{1}{2\pi}\int dq q^{2\alpha+1} [L_n(q^2/2)]^2L_m(q^2)e^{-(1+\frac{1}{2\sigma^2})q^2}. 
\label{eq:3particle}
\end{aligned}
\end{equation}
The three-particle matrix elements in Eq.~\ref{eq:3particle} are linear in the coefficients $c_\alpha$.   We determine these coefficients by minimizing weighted squared deviations of the three-particle spectrum from the target $U_3=1$ and $U_{L>3,i}=0$ (see SM \cite{SM}).
Figure \ref{fig:engineered_interaction}(a) shows the resulting Haldane pseudopotentials, while Fig. \ref{fig:engineered_interaction}(b) shows the corresponding three-particle spectrum. As intended, $U_3$ remains well separated while variations among the higher $L$ energies are suppressed. When evaluated using Eq.~\eqref{eq: cost_function} with the same cutoff and weights as in Fig.~1, the engineered interaction has $\mathcal{L}=0.0445$, smaller than the cost $\mathcal{L}=0.0751$ for $n=1$ Coulomb.

\begin{figure}
    \centering   
\includegraphics[width=0.49\textwidth]{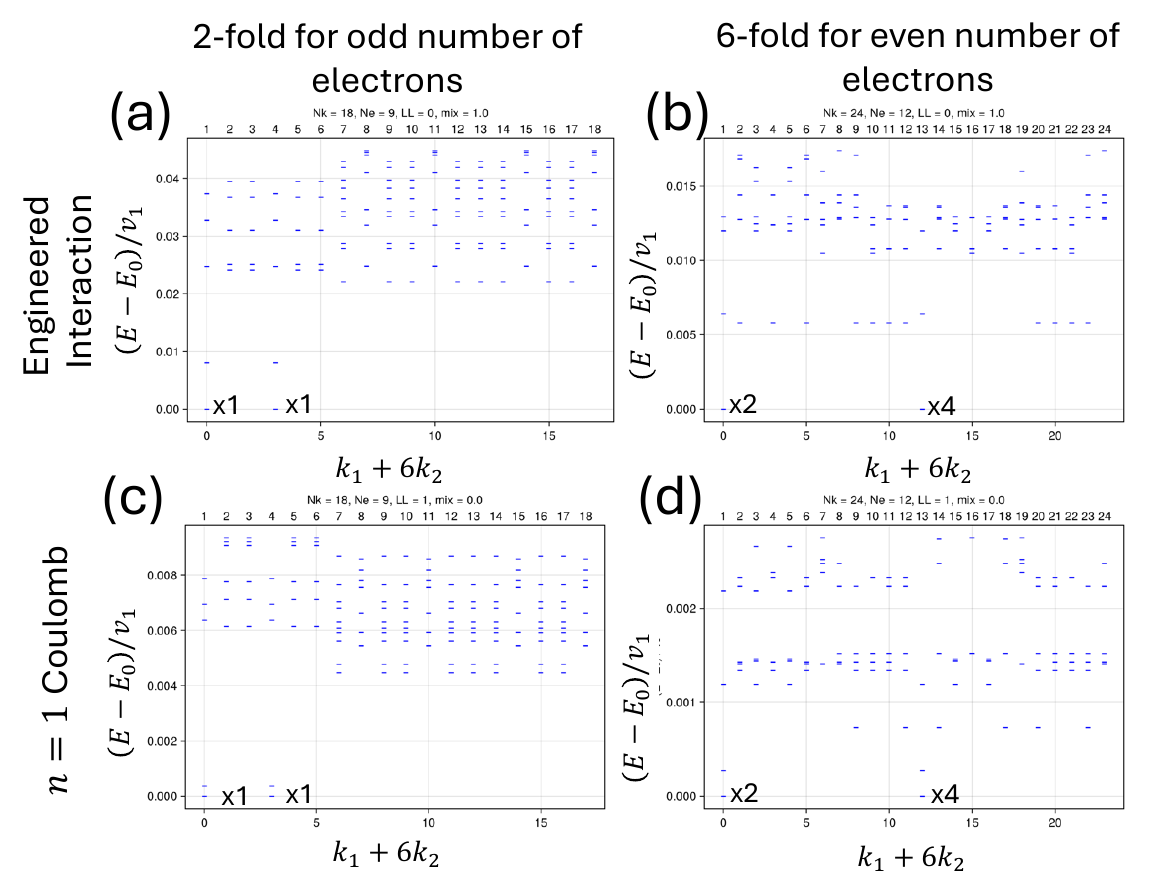}
    \caption{Many-body exact diagonalization spectra for the optimized interaction (a-b) 
     and the Coulomb interaction projected to the $n=1$ LL (c-d) for an odd electron number $(N_e=9)$ (a,c) and an even electron number $N_e=12$ (b,d). The labels indicate the multiplicities of nearly degenerate states. The Moore-Read optimized interaction exhibits the expected ground state 
     quasi-degeneracy pattern: twofold degeneracy for an odd number of electrons, and sixfold degeneracy for an even number of electrons.  The low-energy states 
     appear in the same momentum sectors as for the $n=1$ Coulomb interaction. 
     }
    \label{fig:ED}
\end{figure}

We next test the engineered interaction using many-body exact diagonalization on a torus. Figure \ref{fig:ED} compares the engineered and $n=1$ Coulomb interactions for $N_e=9$ and $12$ electrons. The engineered interaction consistently exhibits the expected Moore–Read even–odd ground-state degeneracy, with the low-energy states occurring in the same momentum sectors as for $n=1$ Coulomb. Additional system sizes are shown in the SM \cite{SM}. The normalized gap separating the ground-state manifold from excited states is larger than for $n=1$ Coulomb for all system sizes studied. 

\textit{Generalizing the Moore–Read criterion to Chern bands---}
The three-particle spectral criterion introduced for LLs above can be readily generalized to Chern bands. 
In the Chern case, the projected interaction is shaped by the band wave functions through their form factors, 
allowing the quantum geometry of the band to modify the three-particle spectrum. 
This spectral approach parallels the use of two-particle exact-diagonalization spectra to identify Haldane-pseudopotential-like structures in Chern bands \cite{lauchli2013hierarchy,ahn2024non}.
In the LL limit, two-particle states appear in pairs related to center-of-mass motion and 
three-particle states appear in triplets. 
We use this structure to group Chern-band three-particle levels and define a generalized Moore–Read cost:
\begin{equation}
\begin{aligned}
    \mathcal{L}_C&=\frac{1}{\mathcal{N}}\sum_{K}
    \sum_{L>3}^{L_{max}} \sum_{i=1}^3 w_L \left(\frac{U_L^i(K)-\bar{U}}{U_3^i(K)-\bar{U}}\right)^2.
\end{aligned}
\label{cost_function_Chern_Band}
\end{equation}
Here, $K$ labels the momentum sector and $i=1,2,3$ labels the three members of each triplet in descending order, $\bar{U}$ is the weighted average over higher $L$ levels $U_L^i(K)$, and the normalization $\mathcal{N}=3 N_K \sum\nolimits_{L>3}^{L_{\max}} w_L $ with $N_K$ the number of momentum sectors.
We assign the three largest eigenvalues in each $K$ sector to the $L=3$ triplet and group the remaining eigenvalues into successive triplets in descending energy order.  Since relative angular momentum
is not a good quantum number in a Chern band, the splitting within an $L$ sector is finite and 
for larger $L$ can be larger than the separation between different $L$ channels.  The 
assignment of 3-particle eigenstates to a nominal $L$ sector is therefore ambiguous for 
larger $L$ channels.  We therefore take $w_L=1$ throughout this analysis to eliminate 
any dependence on these assignments.

\begin{figure}
    \centering   
\includegraphics[width=0.49\textwidth]{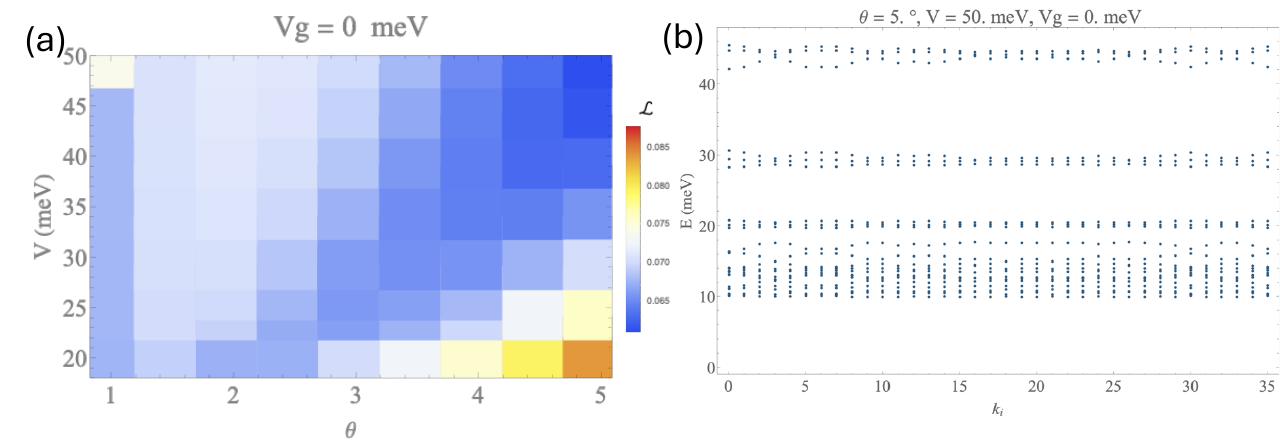}
    \caption{(a) Generalized Moore--Read cost $\mathcal{L}_C$ as a function of twist angle $\theta$ and moiré potential strength $V$ at $V_g=0$. (b) Three-particle spectrum as a function of total momentum sector at $\theta=5^\circ$, $V=50$ meV, and $V_g=0$, corresponding to the lowest-cost point in (a).}  
    \label{fig:tmote2}
\end{figure}

We apply our generalized criterion to twisted MoTe$_2$ moir\'e materials, 
motivated by the valley polarization of its minibands observed \cite{tmote2_review} when the the topmost valence band is partially filled.  Although this band is understood to resemble the $n=0$ LL \cite{AdiabaticAproximation1,AdiabaticAproximation2}, 
our criterion does not rely on proximity to a particular LL. 
We evaluate the generalized cost using the continuum model of twisted MoTe$_2$ \cite{FengchengTopology}, varying the twist angle $\theta\in[1,5]^\circ$, moiré potential strength $V\in[20,50]$ meV, and interlayer potential difference $V_g\in[0,20]$meV, while fixing the interlayer tunneling $w=-30$ meV, moiré potential phase $\psi=120^\circ$, and screening gate distance $d_g=20\,\mathrm{nm}$. 
Figure~\ref{fig:tmote2} shows the cost evaluated from the three-particle spectrum
as a function of twist-angle $\theta$ and the continuum-model \cite{FengchengTopology} potential-strength parameter $V$ at gate potential $V_g=0$.
The low-cost region forms a line in the ($\theta$, $V$) plane qualitatively consistent with the magic-line behavior associated with nearly ideal LL-like bands \cite{LiangFuMagic,AdiabaticAproximation1, nicolas_pressure}. 
Figure~\ref{fig:tmote2}(b) shows the three-particle spectrum at the lowest-cost point in the $V_g=0$ scan, $\theta=5^\circ$, $V=50$ meV. Unlike in LLs, the three-particle levels exhibit splitting in the triplet structure and center-of-mass-momentum dependence
\footnote{
Three-fold degeneracy also occurs in the atomic limit of triangular lattice Hubbard models, but in that case only for the highest energy.}. The highest-energy triplet is well separated from the remaining levels, but the latter still exhibit considerable energy variations rather than forming the nearly degenerate manifold of the ideal Moore–Read parent spectrum.

To test whether these favorable three-particle spectra are accompanied by Moore–Read order, we perform many-body ED calculations at the optimized parameters. While for a $5\times4$ torus, we find a well-separated six-state low-energy manifold, its momentum-sector structure does not fully reproduce the expected Moore–Read pattern.The expected degeneracy structure is also absent in the $4 \times 4$ calculation \cite{SM}. Our results therefore do not provide compelling evidence for Moore–Read order in the parameter regimes studied, highlighting the challenges of stabilizing this phase in twisted MoTe$_2$.

\textit{Discussion --} In this work, we have introduced a three-particle spectral criterion for assessing 
how closely band-projected interacting-electron Hamiltonians resemble the Moore--Read 
three-body parent Hamiltonian.
In LLs, we obtained an analytic mapping between two-body Haldane pseudopotentials and the three-particle spectrum, and proposed a cost function to quantify proximity to Moore-Read. 
Using the parent spectrum as a target, we designed two-body interactions 
with enhanced Moore--Read physics signatures. 
We then generalized the cost function to Chern bands and applied it to twisted MoTe$_2$. 
Despite identifying regions with relatively low cost, our many-body calculations do not provide 
compelling evidence for Moore--Read order in the regimes studied.  

Our finding that it is not easy to stabilize Moore--Read states speaks to the subtlety, 
even in Landau levels, of its competition with other correlated phases. At half filling, Coulomb interactions in the $n=0$, $n=1$, and $n=2$ LLs favor composite Fermi liquid, non-Abelian paired, and charge-density-wave states, respectively. Despite the qualitative differences between these states, the 
Haldane pseudopotentials that select them differ only moderately in their short-range pseudopotential ratios $(v_3/v_1,v_5/v_1)$  (see Fig.~\ref{fig:CoulombLLs} and Fig.~\ref{fig:schematic_phase_diagram} in the SM \cite{SM}). This suggests a picture in which Moore--Read pairing emerges in an intermediate region of the model electron-electron interaction phase space - between competing composite Fermi liquid and charge-ordered phases, consistent with studies of conventional LLs \cite{rezayi2000incompressible} and moiré Chern bands \cite{emil_nonAbelian}.
The delicacy of the competition between very distinct states in LLs provides a useful perspective on the challenge of stabilizing Moore--Read order in other systems, such as twisted MoTe$_2$.

A translationally invariant two-body interaction projected to a single LL respects particle-hole symmetry \cite{allan_QH_notes}. As with the Coulomb interaction projected to the $n=1$ LL, our engineered interaction therefore does not explicitly distinguish the Pfaffian from its particle-hole conjugate, the anti-Pfaffian \cite{apf1,apf2}. LL mixing, however, breaks this symmetry by generating effective three-body interactions, in addition to modifying the two-body pseudopotentials \cite{bishara2009effect,inti_LL_mixing}. A natural extension of our framework is to apply the three-particle spectral criterion to these effective interactions, and use it to investigate how LL mixing influences the stability of Moore--Read order. Bilayer graphene provides an attractive platform for exploring these effects because of its electrically tunable LL structure. Band mixing in twisted MoTe$_2$ can generate similar effective higher-body interactions, and explicit three-body interactions have been proposed as a route to stabilizing the Moore--Read state \cite{zhang2024moore}. We leave these interesting directions for future work. 

\textit{Acknowledgments --} BMK acknowledges helpful discussions with Steven Simon, Patrick Ledwith, Ahmed Abouelkomsan, and Xiaohan Wan. Numerical exact diagonlization calculations were
performed using MomentumED.jl libraries. This work was supported by a grant from the Simons Foundation [MPS-T-MPS-00884934,AM].
Y.-C.C. and J.-J.S. acknowledges support from the
National Science and Technology Council (NSTC) of Taiwan
under Grant No. 114-2112-M-A49-031 and 115-2112-M-A49-025 and the Top Research Centers in Taiwan Key Fields Program, funded by the Ministry of Education (MOE), Taiwan.
\bibliography{refs}

\newpage

\onecolumngrid
\appendix
\section{Supplemental Material}
\subsection{Three-particle basis and mapping from two-body interactions}
We use a basis for the three-body fermionic relative-motion Hilbert space derived by Laughlin \cite{laughlin_3body}:
\begin{equation}
    \ket{\mu,\nu}
    =
    \frac{1}{[2^{6\mu+4\nu+1}(3\mu+\nu)!\nu!\pi^2]^{1/2}}
    \left[
    \frac{(z_a+i z_b)^{3\mu}-(z_a-i z_b)^{3\mu}}{2i}
    \right]
    (z_a^2+z_b^2)^\nu
    e^{-(|z_a|^2+|z_b|^2)/4},
    \label{basis_SM}
\end{equation}
where $\mu=1,2,\ldots$, $\nu=0,1,2,\ldots$, and  
\begin{equation}
    \begin{aligned}
        Z&=(z_1+z_2+z_3)/3,\\
        z_a&=\sqrt{\frac{2}{3}}
        \left(\frac{z_1+z_2}{2}-z_3\right),\\
        z_b&=\frac{1}{\sqrt{2}}(z_1-z_2),
    \end{aligned}
    \label{jacobi_coordinates_SM}
\end{equation}
are the center-of-mass (COM) and relative Jacobi coordinates.
We now show explicitly that the states in Eq.~\eqref{basis_SM} span the antisymmetric three-particle relative-motion Hilbert space. A general LLL relative motion polynomial 
can be written as 
\begin{equation}
    f(z_+,z_-)= \sum_{p,q\geq0} c_{p,q}z_+^pz_-^q.
\end{equation}
where $z_\pm= (z_a \pm i z_b)/\sqrt{2}$. Antisymmetry can be enforced by imposing invariance under cyclic permutation of the three particle coordinates $1 \to 2 \to 3 \to 1$ and antisymmetry under the exchange of $1\leftrightarrow2$. Under $1 \leftrightarrow2$, $z_a \to z_a,~ z_b\to-z_b$ and so $z_\pm \to  z_\mp$. Antisymmetry then requires  
\begin{equation}
    f(z_+,z_-)= \frac12 \sum_{p,q\geq0} c_{p,q}(z_+^pz_-^q-z_+^qz_-^p)
\end{equation}
which vanishes for $p=q$.  Interchanging $p,q$ gives an overall minus sign only, rather than generating independent terms so we can restrict the sum to
\begin{equation}
    f(z_+,z_-)= \sum_{p>q\geq0} c_{p,q}(z_+^pz_-^q-z_+^qz_-^p)
\end{equation}
Cyclic permutation produces a $120^\circ$ rotation in the $z_a, z_b$ plane so that
$z_\pm \to e^{\pm i 2\pi/3}z_\pm$ and  
\begin{equation}
    f(z_+,z_-) \to \sum_{p>q\geq0} c_{p,q}(z_+^p z_-^q e^{2\pi i (p-q)/3}-z_+^qz_-^p e^{-2\pi i (p-q)/3}).
\end{equation}
It follows that only terms with $p-q=3\mu, \mu=1,2,..$ are allowed.
We can then define $\nu=q, p=3\mu+\nu$.  The general antisymmetric relative-motion polynomial then becomes
\begin{equation}
    f(z_+,z_-)=\sum_{\mu\geq1,\nu\geq0}c_{\mu\nu}
(z_+^{3\mu}-z_-^{3\mu})(z_+z_-)^\nu.
\end{equation}
Individual terms in this sum are equivalent to the basis functions in Eq. \eqref{basis_SM},
up to normalization constants. Orthogonality follows from the orthogonality of distinct monomials, with the cross terms vanishing since they require $\mu+\mu'=0$, which is impossible for $\mu,\mu'\geq1$.  

We now derive the matrix elements of a two-body interaction in the three-particle basis. Consider the  interaction $V_{12}$ between particles 1 and 2, parametrized by its Haldane pseudopotentials,
\begin{equation}
    V_{12}=\sum_{M,m} v_m \ket{M,m}_{12} \bra{M,m}_{12}
    \label{eq: Haldane_expansion}
\end{equation}
where $M$ and $m$ label the pair center-of-mass and relative angular momenta respectively. In the three particle Hilbert space, the interaction acts an identity on particle 3, 
\begin{equation}
    V_{12}=\sum_{M,m} v_m \ket{M,m}_{12} \bra{M,m}_{12}\otimes \mathbf{1_3}=\sum_{M,m,l} v_m \ket{M,m;l}_{123} \bra{M,m;l}_{123}
    \label{eq: Haldane_expansion_13}
\end{equation}
where $    \ket{M,m;l}=\ket{M,m}_{12}\otimes \ket{l}_3$ and $\ket{l}_3$. is a single-particle LLL orbital. In coordinate representation, these states have the polynomial structure
\begin{equation}
    \langle z_1,z_2,z_3 | M,m;l \rangle \sim (z_1+z_2)^M(z_1-z_2)^m z_3^l.
    \label{eq: Mml zi}
\end{equation}
Directly evaluating the matrix elements in this basis is inconvenient because the pair COM and third-particle coordinates mix the three-particle COM coordinate $Z$ with the relative coordinate $z_a$,
\begin{equation}
     (z_1+z_2)^M(z_1-z_2)^m z_3^l\sim 
    (2 Z+\sqrt{\frac{2}{3}} z_a)^M(z_b)^m (Z-\sqrt{\frac{2}{3}}z_a)^l
    \label{eq:2_body_in_jacobi}
\end{equation}
We can avoid this by changing to a basis adapted to the Jacobi coordinates. 
From Eq.~\eqref{eq:2_body_in_jacobi}, we see that the pair COM coordinate $z_1+z_2$ and the third-particle coordinate $z_3$ can be expressed in terms of the three-particle COM coordinate $Z$ and the relative coordinate $z_a$, without affecting $z_b$. This allows us to make a change of basis from those proportional to $ (z_1+z_2)^M z_3^l$ to $(z_1+z_2+z_3)^M z_a^A$. We define the latter as 
\begin{equation}
    \langle Z, z_a,z_b|MAm\rangle= \mathcal{N}_{MAm} Z^M z_a^A z_b^m e^{-(3 |Z|^2+|z_a|^2+|z_b|^2)/4}
\end{equation}
where $\mathcal{N}_{MAm}$ is the normalization factor. 
Since the two sets of states form complete bases of the same fixed $m$ subspace,
we can equivalently write
\begin{equation}
\sum_{M,l}\ket{M,m;l}\bra{M,m;l}
=
\sum_{M,A}\ket{M,A,m}\bra{M,A,m},
\end{equation}
and the interaction becomes
\begin{equation}
     V_{12}= \sum_{m}v_m \sum_{M,A} \ket{M,A,m}\bra{M,A,m}
\label{eq:Haldane_expansion_MAm}
\end{equation}
The normalization constant $\mathcal{N}_{MAm}$ can be determined using the Jacobi-coordinate transformation, 
\begin{equation}
\begin{aligned}
\sum_{i=1}^3\left|z_i\right|^2 & =3|Z|^2+\left|z_a\right|^2+\left|z_b\right|^2, \\
d^2 z_1 d^2 z_2 d^2 z_3 & =3 d^2 Z d^2 z_a d^2 z_b ,
\end{aligned}
\end{equation}
which factorizes to 
\begin{equation}
    \mathcal{N}_{MAm}= [\frac{1}{2 \pi 2^m m!}\cdot \frac{1}{2\pi 2^A A!}\cdot \frac{1}{2 \pi (2/3)^MM!}]^{1/2}
\end{equation}
Here the COM normalization incorporates the factor of three in the integration measure. 

With the interaction expressed in this basis, our remaining task is to evaluate the overlaps $\langle M,A,m|\lambda,\mu,\nu\rangle$, where $\lambda$ labels the three-particle COM angular momentum. This amounts to expanding the polynomial part of Eq. \eqref{basis_SM} in powers of $z_a$ and $z_b$. Using the binomial expansion,
\begin{equation}
\begin{aligned}
\frac{(z_a+iz_b)^{3\mu}-(z_a-iz_b)^{3\mu}}{2i}
(z_a^2+z_b^2)^\nu=
\sum_{k=0}^{\lfloor(3\mu-1)/2\rfloor}
\sum_{j=0}^{\nu}
(-1)^k
\binom{3\mu}{2k+1}\binom{\nu}{j}
z_a^{L-(2k+2j+1)}
z_b^{2k+2j+1},
\end{aligned}
\end{equation}
where $L=3\mu+2\nu$. Defining $m= 2k+2j+1$, the sum can be written as 
\begin{equation}
    \frac{(z_a+iz_b)^{3\mu}-(z_a-iz_b)^{3\mu}}{2i}
(z_a^2+z_b^2)^\nu=\sum_{\substack{1 \leq m \leq L \\ m \text { odd }}} C_{\mu \nu}^m z_a^{L-m} z_b^m,
\end{equation}
where
\begin{equation}
    C_{\mu \nu}^m=\sum_{j=0}^\nu(-1)^{\frac{m-1}{2}-j}\binom{3 \mu}{m-2 j}\binom{\nu}{j} .
\end{equation}
Including the normalization factors, the overlap
\begin{equation}
\langle M, A, m \mid \lambda, \mu, \nu\rangle=\delta_{M, \lambda} \delta_{A, L-m} \chi_{\mu \nu}^m,
\end{equation}
where 
\begin{equation}
    \chi_{\mu \nu}^m=\left[\frac{m!(L-m)!}{2^{L-1}(3 \mu+\nu)!\nu!}\right]^{1 / 2} C_{\mu \nu}^m .
\end{equation}
Substituting the overlaps into the pseudopotential expansion of \(V_{12}\) gives
\begin{equation}
\begin{aligned}
\langle\lambda',\mu',\nu'|V_{12}|
\lambda,\mu,\nu\rangle
&=
\sum_m v_m\sum_{M,A}
\langle\lambda',\mu',\nu'|M,A,m\rangle
\langle M,A,m|\lambda,\mu,\nu\rangle
=
\delta_{\lambda,\lambda'}\delta_{L,L'}
\sum_m v_m
\chi_{\mu'\nu'}^{m}\chi_{\mu\nu}^{m}.
\label{eq:mapping_SM}
\end{aligned}
\end{equation}
We therefore obtain the mapping between the two-body pseudopotentials and the three-particle interaction matrix elements,
\begin{equation}
P_m(\mu,\nu,\mu',\nu')
=
\delta_{L,L'}
\chi_{\mu\nu}^{m}\chi_{\mu'\nu'}^{m}
\end{equation}
which is Eq. (7) in the main text. For one dimensional relative angular momentum sectors, $L=L'$ implies $\mu=\mu'$ and $\nu=\nu'$, so the three-particle energy is
\begin{equation}
U_L=\sum_m v_m|\chi_{\mu\nu}^{m}|^2.
\end{equation}
For multi-dimensional sectors, the energies $U_{L,i}$ are obtained by diagonalizing their corresponding matrix. The normalization of the basis provides a consistency check:
\begin{equation}
1
=\langle \lambda , \mu,\nu|\lambda, \mu,\nu\rangle
=
    \sum_{MAm}\langle \lambda \mu \nu |MAm\rangle\langle MAm |\lambda\mu\nu\rangle
=\sum_{\substack{1\leq m\leq L\\m\ \mathrm{odd}}}
|\chi_{\mu\nu}^{m}|^2,
\end{equation}
which we verified numerically and is consistent with the understanding of $|\chi_{\mu\nu}^{m}|^2$ being the probability that a given pair occupies relative-angular-momentum channel $m$. As an additional check, we compare the three-particle spectrum obtained from the analytical mapping with direct three-particle exact diagonalization, as discussed below.
Since the matrix elements of $V_{12}, V_{13}$, and $V_{23}$ are identical between fully antisymmetric three-particle states, the corresponding full-triplet interaction has matrix elements three times those of $V_{12}$.  The $U_{L,i}$ values in the main text are in units of energy per pair.
\subsection{Additional Analysis of the Cost function}
\begin{figure}
    \centering   
\includegraphics[width=1\textwidth]{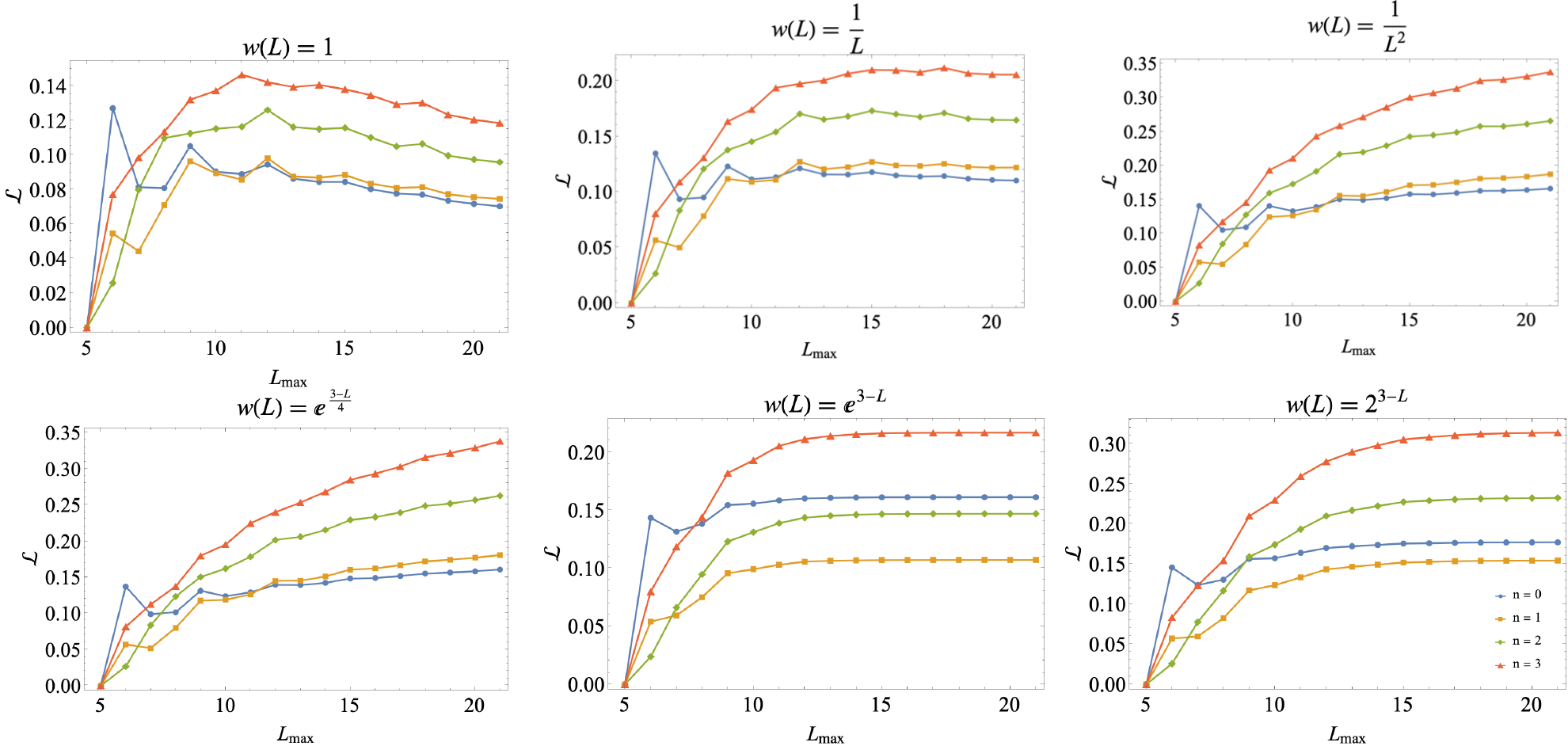}
    \caption{Cutoff and weight dependence of the cost function for Coulomb interactions projected to different LLs }  
    \label{fig:cutoffs}
\end{figure}

\begin{figure}
    \centering   
\includegraphics[width=0.5\textwidth]{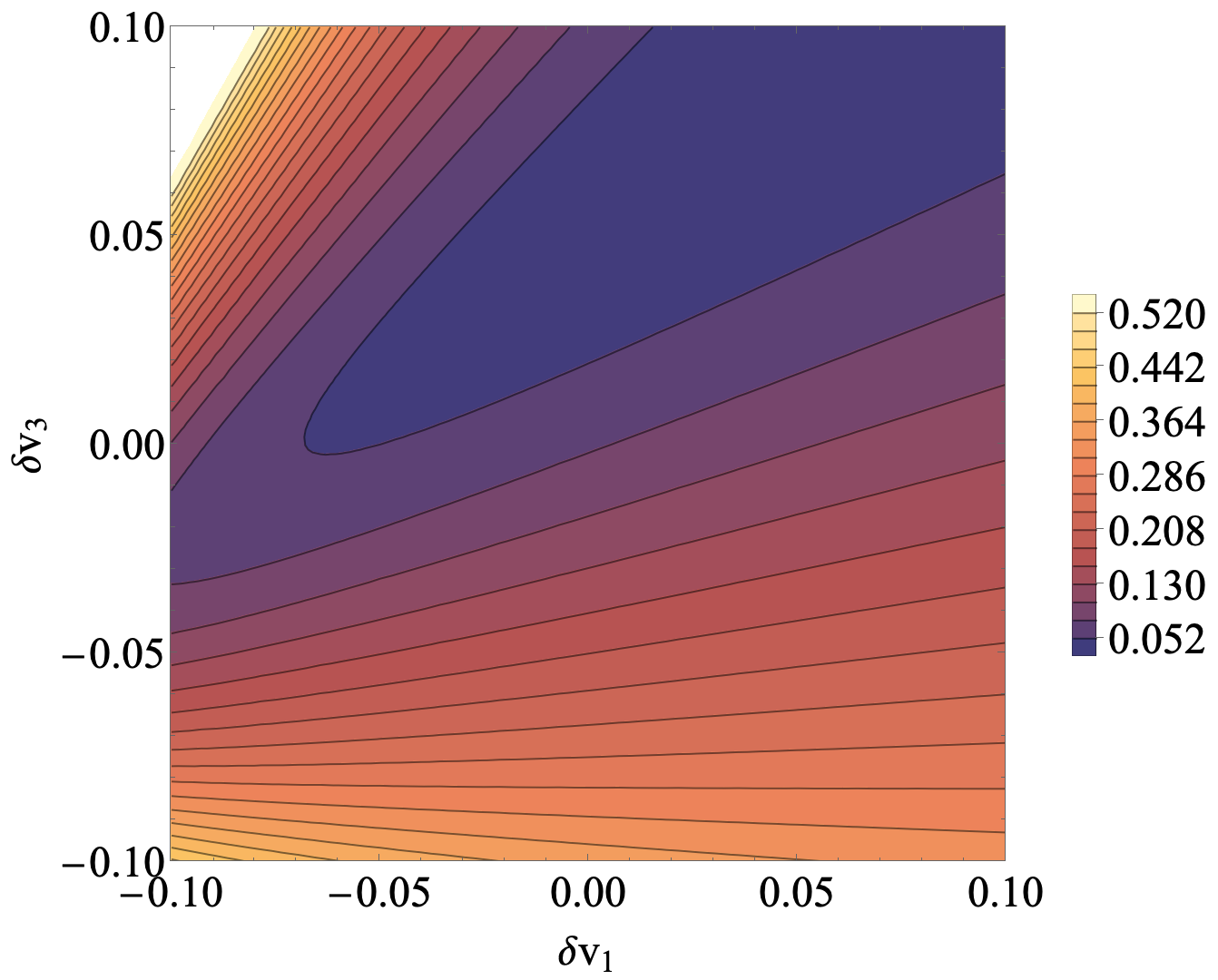}
    \caption{Moore--Read cost $\mathcal{L}$ as a function of variations $\delta v_1$ and $\delta v_3$ from their $n=1$ Coulomb values, with higher pseudopotentials held fixed. The cost is evaluated using $w_L=e^{-(L-3)}$ and $L_{\max}=8$. }  
    \label{fig:cost_contour_n=1}
\end{figure}

The weights $w_L$ and angular-momentum cutoff $L_{\max}$ entering the cost function in Eq.~\eqref{eq: SM_cost_function} are not uniquely determined. To examine the sensitivity of our conclusions
to these choices, we evaluate the cost for Coulomb interactions projected onto the $n=0,1,2,3$ Landau levels using several weighting schemes for different cutoffs $L_{\max}$, as illustrated in Fig.~\ref{fig:cutoffs}. We find that the $n=1$ LL has the lowest cost over a range of cutoffs for all weighting schemes considered, although the extent of this range depends on the choice of weights. We also examine the dependence of the cost function on variations of the two-body pseudopotentials $v_1$ and $v_3$ about their $n=1$ Coulomb values, keeping the remaining pseudopotentials fixed. The resulting cost landscape, shown in Fig.~\ref{fig:cost_contour_n=1}, exhibits an extended low-cost region whose orientation is qualitatively consistent with the ridge of enhanced Moore--Read overlap and excitation gap found in previous many-body exact-diagonalization studies \cite{peterson2008orbital,storni2010fractional}.
\subsection{Inverse-design optimization}
The ground state of any 3-body interaction projected to a single Landau level
is independently invariant under rigid shifts and under positive rescaling of the three particle spectrum.
In the main text, we therefore compare interactions using the scale and 
shift invariant cost function,
\begin{equation}
    \mathcal{L}= \frac{1}{\mathcal{N}}\sum_{L>3,i}^{L_{max}} w_L (\frac{U_{L,i}-\bar{U}}{U_3-\bar{U}})^2
\label{eq: SM_cost_function}
\end{equation}
where the normalization factor 
$\mathcal{N}=\sum\nolimits_{L>3}^{L_{\max}} w_L d_L$,
$i$ labels eigenenergies within a degenerate $L$ sector, $w_L$ is the weight assigned to 
angular momentum channel $L$, 
and $\bar{U}$ is the normalized weighted average of $U_{L,i}$ over $L >3$. 
To estimate optimal pair-wise interactions it is convenient to 
instead use the surrogate quadratic objective,
\begin{equation}
\mathcal{L}_{\rm sur}
=
w_3(U_3-1)^2+
\sum_{L>3,i}w_L(U^{(L)}_i)^2,
\label{eq:surrogate_cost}
\end{equation}
with a large value of $w_3$ to impose the scale $U_3 \approx 1$.  Since the three-particle matrix elements 
are linear in the coefficients $c_\alpha$, $\mathcal{L}_{\rm sur}$ is quadratic 
giving a weighted linear least-squares problem in $c_\alpha$. 

In Eq. \eqref{eq:surrogate_cost} we first note that within each $L$ sector,
\begin{equation}
    \sum_i
U_{L,i}^2= \mathrm{Tr}\left [ (U^{(L)})^2\right ] = 
\sum_{a,b}U^{(L)}_{a,b} U^{(L)}_{b,a}=\sum_{a,b} (U_{a,b}^{(L)})^2=\sum_{a}(U^{(L)}_{a,a})^2+2 \sum_{a<b} (U^{(L)}_{a,b})^2, 
\label{eq:evals_trace_identity}
\end{equation}
which allows us to express the cost \eqref{eq:surrogate_cost} in terms of the matrix elements rather than the eigenvalues.  Here $a$ and $b$ are lumped $(\mu,\nu)$ three-particle state labels.
It follows that 
\begin{equation}
    \mathcal{L}_{\rm sur}
=
w_3(U_3-1)^2+
\sum_{L>3,i}w_L(U^{(L)}_i)^2=w_3(U_3-1)^2+\sum_{L>3,a,b} w_L (U^{(L)}_{a,b})^2
\end{equation}
Treating the one-dimensional case ($L \le 8$) first, we note that $U_L$ depends linearly 
on the coefficients $c_{\alpha}$ used to specify the pair interaction:
\begin{equation}
    U_L=\sum_{\alpha}T_\alpha (\mu_L,\nu_L, \mu_L,\nu_L) c_\alpha = \sum_{\alpha}T_{L,\alpha} c_\alpha 
\end{equation}
where $(\mu_L,\nu_L)$ are the labels corresponding to the angular momentum $L$, and for convenience we have defined $T_\alpha (\mu_L,\nu_L, \mu_L,\nu_L)\equiv T_{L,\alpha} $. 
Differentiating the cost function and setting it to zero,
\begin{equation}
\begin{aligned}
    \frac{\partial{\mathcal{L}_{\rm sur}}}{\partial c_\beta}&= 2 w_3 (U_3-1)  \frac{\partial U_3}{\partial c_\beta}+2 \sum_{L>3}w_L U_L  \frac{\partial U_L}{\partial c_\beta} \\
    &=2 w_3 ((\sum_\alpha T_{3,\alpha} c_\alpha) -1)  T_{3,\beta}+2 \sum_{L>3}w_L (\sum_\alpha T_{L,\alpha} c_\alpha )  T_{L,\beta}=0.
\end{aligned}
\end{equation}
Simplifying, 
\begin{equation}
          w_3T_{3,\beta}=\sum_\alpha  \left( w_3 T_{3,\alpha}  T_{3,\beta} + \sum_{L>3} w_L  T_{L,\alpha}  T_{L,\beta}\right) c_\alpha 
\end{equation}
which can be solved for the $c_{\alpha}$ by matrix inversion.

For the case of non-unit dimension $L$ sectors, let these states within a degenerate sector $L$ with dimension $d_L$ be labeled by $(\mu_a, \nu_a), a=1,..,d_L$ and the define the matrix
\begin{equation}
    [\mathbf{T}_\alpha^{(L)}]_{a,b}=T_\alpha(\mu_a,\nu_a, \mu_b, \nu_b)
\end{equation}
with the corresponding matrix
\begin{equation}
    \mathbf{U}^{(L)}=\sum_\alpha c_\alpha \mathbf{T}_\alpha^{(L)}
    \label{eq:U=Tc}
\end{equation}
and writing the surrogate cost in terms of the matrix instead of the eigenvalues using the identity  Eq. \eqref{eq:evals_trace_identity}, 
\begin{equation}
    \mathcal{L}_{\mathrm{sur}}= w_3 (U_3-1)^2+\sum_{L>3} w_L \mathrm{Tr}[(\mathbf{U}^{(L)})^2].
\label{eq:matrix trace cost}
\end{equation}
Differentiating the higher $L$ terms ,
\begin{equation}
    \frac{\partial}{\partial c_\beta}\mathrm{Tr}[(\mathbf{U}^{(L)})^2]=2 \mathrm{Tr}[\mathbf{U}^{(L)}\frac{\partial\mathbf{U}^{(L)}}{\partial c_\beta}]
\end{equation}
And from Eq. \eqref{eq:U=Tc},
\begin{equation}
    \frac{\partial\mathbf{U}^{(L)}}{\partial c_\beta}=\mathbf{T}_\beta^{(L)}
\end{equation}
Again, differentiating Eq. \eqref{eq:matrix trace cost} and setting the derivative to zero, 
\begin{equation}
    w_3 T_{3,\beta}= \sum_\alpha\left ( w_3T_{3,\alpha} T_{3,\beta} + \sum_{L>3} w_L \mathrm{Tr [T_\alpha^{(L)} T_\beta^{(L)} ]}\right)c_\alpha
\end{equation}
which again is a matrix inverse problem.

\begin{figure}
    \centering   
\includegraphics[width=0.55\textwidth]{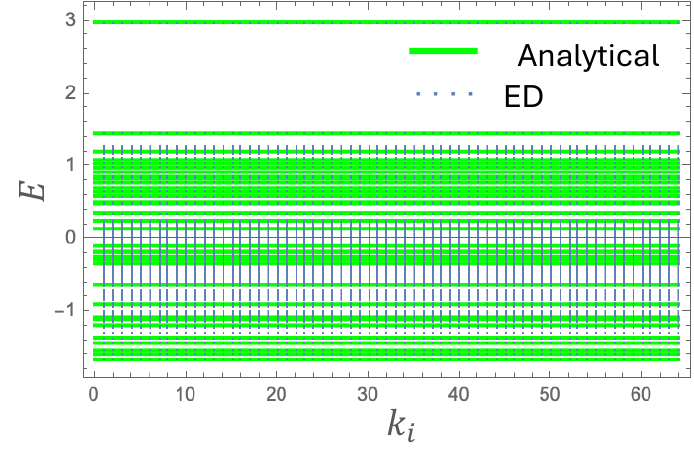}
    \caption{Comparison of the analytical three body spectrum for angular momentum up to $L_{max}=21$ obtained from the mapping in Eq.~\eqref{eq:mapping_SM} to the numerical 3-body ED spectrum for the same interaction in a finite toriodal system with 64 states per Landau level.  These results show that finite-size corrections are small up to $L_{max}$.
    The degeneracy with respect to the momentum label $k_i$ originates from the center-of-mass degree-of-freedom.  The ED spectra include some states with relative angular momentum larger than
    $L_{max}$.} 
    \label{fig:SM_comparison}
\end{figure}

This surrogate cost can be understood as trying to target the spectrum $U_3=1$ and $U_{L>3,i}=0$ as closely as possible in a least squares sense. To see this explicitly, we define $A$ using the independent matrix elements $a\leq b$,
\begin{equation}
    A_{(L,a,b),\alpha}\equiv T_\alpha(\mu_a,\nu_a,\mu_b,\nu_b),
\end{equation}
the target vector,
\begin{equation}
    b_{(L, a, b)}= \begin{cases}1, & L=3, \\ 0, & L>3 .\end{cases}
\end{equation}
and the diagonal weight matrix
\begin{equation}
W_{(L, a b),(L, a, b)}= \begin{cases}w_3, & L=3, \\ w_L, & L>3, a=b, \\ 2 w_L, & L>3, a<b .\end{cases}
\end{equation}
where the factor of two is to account for summing over half of the off-diagonal elements. The cost function $\mathcal{L}_{\mathrm{sur}}$ can then be written as a weighted least squares:
\begin{equation}
\mathcal{L}_{\mathrm{sur}}  
=
(A \mathbf{c}-\mathbf{b})^T W( A \mathbf{c} -\mathbf{b})
\end{equation}
This makes explicit that the optimization minimizes the weighted deviations of the three-particle matrix elements from those of the Moore–Read parent Hamiltonian.
For each $\sigma$ we minimize $\mathcal{L}_{\rm sur}$ over $c_\alpha$
and subsequently optimize over $\sigma$.  The resulting interaction is
then evaluated using $\mathcal{L}$ with the same weights and
angular-momentum cutoff used for the Coulomb Landau levels.  All
reported comparisons therefore use the same shift- and
scale-independent metric. In the optimization shown in the main text we take
$L_{\max}=21$, $\alpha_{\max}=7$, $w_3=10$, and
$w_L=1/L^2$ for $L>3$. We use weak ridge regularization with $\lambda=2.5 \times10^{-10}$ coefficient regularization and scan $\sigma$. The resulting coefficients are $c_\alpha=2\pi \{49.334, -27.3178, -27.4744, 4.18167, 10.7791, -5.27089, 0.707148\}$ and $\sigma=0.632$. 
In Fig.~\ref{fig:SM_comparison}, we compare the analytical three-particle spectrum of the optimized interaction with direct three-particle ED, finding good agreement and providing an independent numerical check of the analytical mapping derived above. For this comparison, the analytical energies are multiplied by three to account for all three interacting pairs, as included in the ED Hamiltonian.
\subsection{Additional Exact Diagonalization Results}
\begin{figure}
    \centering   
\includegraphics[width=1\textwidth]{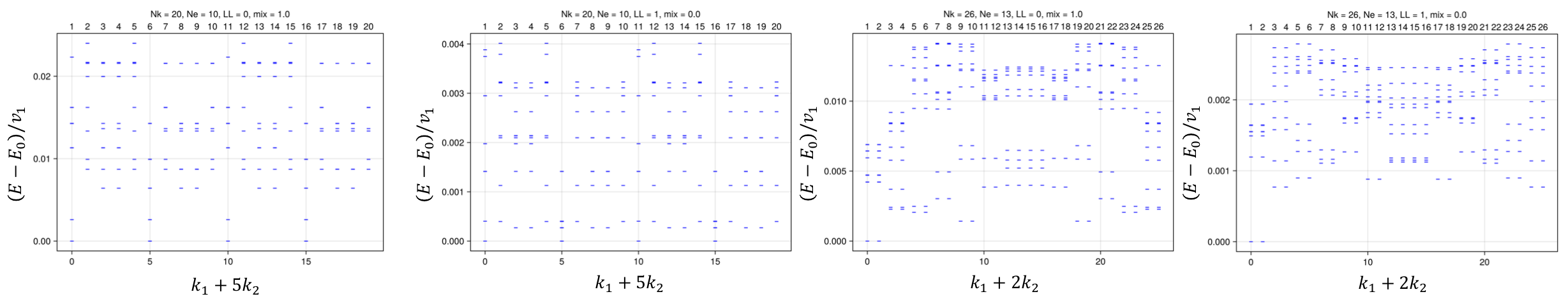}
    \caption{Extended ED data for $N_e=10,13$ for the engineered interaction and $n=1$ Coulomb. } 
    \label{fig:SM_ED}
\end{figure}

To further test the robustness of the many-body ED results presented in the main text, we consider additional system sizes. Figure~\ref{fig:SM_ED} shows the ED spectra for $N_e=10,13$, comparing the engineered interaction with the $n=1$ Coulomb interaction. The engineered interaction exhibits the expected Moore–Read even–odd ground-state degeneracy pattern, with the low-energy states occurring in the same momentum sectors as for $n=1$ Coulomb for all system sizes studied. The normalized gaps $\Delta/v_1$ for the enginnered interaction are larger than those of $n=1$ Coulomb by factors of $\{ 22.1,9.62,21.3,1.85\}$ for $N_e=9,10,12,13$, respectively.

The ED calculations are performed on periodic finite-size clusters of a triangular lattice with
$N_k=18,20,24,26$ momentum points. The allowed momenta are $\mathbf{k}_{mn}=m\mathbf{G}_1/N_1+n\mathbf{G}_2/N_2$, and $(N_1,N_2)=(6,3),(5,4),(6,4), (2,13)$ respectively. The corresponding reciprocal lattice vectors are specified by $(|\mathbf G_2|/|\mathbf G_1|,\cos\theta)=(1,-1/2),(4/5,-1/2),(2/3,-1/2),(\sqrt7,-5/(2\sqrt7) $, respectively, where $\theta$ is the angle between $\mathbf G_1$ and $\mathbf G_2$.

For the exact diagonalization spectra of twisted MoTe$_2$, we use the momentum convention $k_{mn}$ now with
$(N_1,N_2)=(4,4)$ and $(5,4)$. The reciprocal lattice
vectors are
$\bm G_1= [4\pi/\sqrt{3} \,a_M] \, (1,0) $  and  $\bm G_2= [4\pi/\sqrt{3} \,a_M] \ (-{1}/{2}, -{\sqrt{3}}/{2}) $,
where $a_M$ is the moir\'e lattice constant. The system is at half-filling with electron number $N_e=8$ and $10$ for $(N_1,N_2)=(4,4)$ and $(5,4)$, respectively. The same
moir\'e parameters are used as in the main text:
$(m, \theta, \psi, V, w) = (0.6 \, m_e, 5.0^\circ, 120.0^\circ, 50 \,{\rm meV}, -30 \,{\rm meV})$.

\begin{figure}
	\centering   
	\includegraphics[width=.9\textwidth]{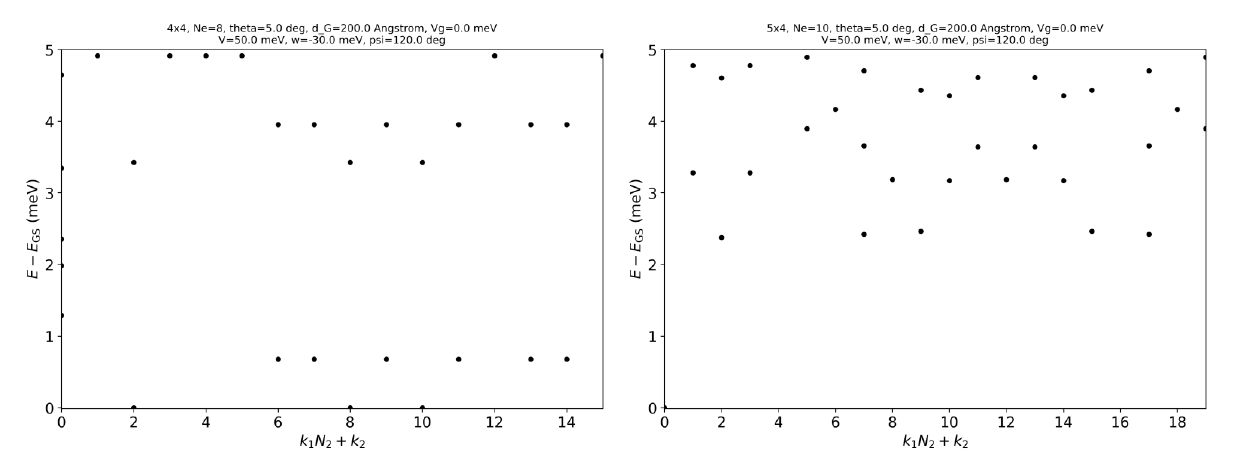}
	\caption{ED spectra for tMoTe$_2$ at half-filling for $(N_1, N_2) = (4,4)$ with $N_e=8$ (left) and $(5,4)$ with $N_e=10$ (right)}  
	\label{fig:ED_tTMD}
\end{figure}

\subsection{Competing Phases at Half Filling}

\begin{figure}
    \centering   
\includegraphics[width=0.5\textwidth]{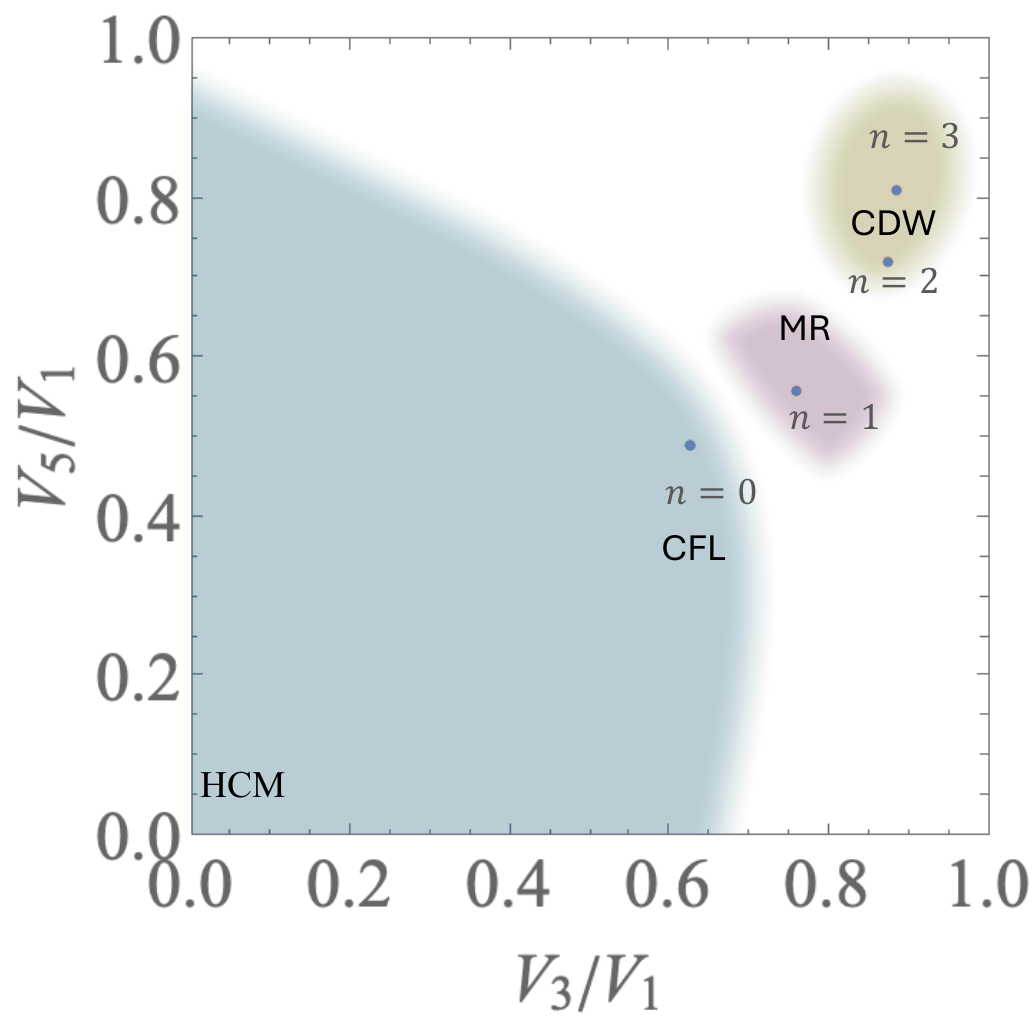}
    \caption{Schematic phase diagram illustrating the competition between composite Fermi liquid (CFL), Moore--Read (MR) paired, and charge-density-wave (CDW) phases in the space of short-range Haldane pseudopotential ratios $(v_3/v_1,v_5/v_1)$. The points indicate Coulomb interactions projected onto the $n=0,1,2,3$ LLs. }  
    \label{fig:schematic_phase_diagram}
\end{figure}
\end{document}